\documentclass[12pt]{article}

\newcommand{\xb}{\mbox{\boldmath $x$}}
\newcommand{\half}{\frac{1}{2}}
\newcommand{\fraction}{\frac{\displaystyle \pi}{\displaystyle 2q}}
\def\sech{\mathop{\rm sech}\nolimits}
\def\csch{\mathop{\rm csch}\nolimits}

\title{
Abrupt termination of a quantum channel and exactly solvable position-dependent mass
models in three dimensions}

\author{C.\ Quesne \thanks{{\sl E-mail address}: cquesne@ulb.ac.be} \\
{\small\sl Physique Nucl\'eaire Th\'eorique et Physique Math\'ematique,  Universit\'e Libre
de Bruxelles,} \\  
{\small \sl Campus de la Plaine CP229, Boulevard~du Triomphe, B-1050
Brussels, Belgium}}
\date{ }
\begin{document}
\baselineskip=22pt plus 1pt minus 1pt
\maketitle

\begin{abstract}
We consider a particle with a position-dependent mass, moving in a three-dimensional
semi-infinite parallelepipedal or cylindrical channel under the influence of some hyperbolic
potential. We show that the lack of uniformity in the environment generates an infinite
number of bound states.
\end{abstract}

\vspace{0.5cm}

\noindent
{\sl PACS}: 03.65.-w

\noindent
{\sl Keywords}: Schr\"odinger equation; Position-dependent mass; Quantum channels
 
\newpage
%
%
\section{Introduction}

Quantum mechanical systems with a position-dependent (effective) mass (PDM) have attracted a
lot of attention and inspired intense research activites during recent years. They are indeed very
useful in the study of many physical problems, such as electronic properties of
semiconductors~\cite{bastard} and quantum dots~\cite{serra}, nuclei~\cite{ring}, quantum
liquids~\cite{arias}, $^3$He clusters~\cite{barranco}, metal clusters~\cite{puente}, etc.\par
%
%
Looking for exact solutions of the Schr\"odinger equation with a PDM has become an interesting
research topic because such solutions may provide a conceptual understanding of some physical
phenomena, as well as a testing ground for some approximation schemes. Although mostly
one-dimensional equations have been considered up to now, several works have recently paid
attention to $d$-dimensional problems~\cite{chen, dong, cq06, mustafa, ju, gonul}.\par
%
%
In \cite{cq06}, we have presented a model depicting a particle moving in a two-dimensional
semi-infinite layer, which may be of interest in the study of quantum wires with an abrupt
termination in an environment that can be modelled by a dependence of the carrier effective mass
on the position.\par
%
%
The purpose of the present Letter is to free ourselves from the restriction to two dimensions
by proposing two novel, more realistic PDM models in three dimensions, which may describe an
abrupt termination of a quantum channel. As we plan to show, these models have also the
advantage of being exactly solvable.\par
%
%
\section{PDM Schr\"odinger equations in three dimensions}

Because of the non-commutativity of the momentum and PDM operators, it is well known that
there is an ordering ambiguity in the kinetic energy operator $T$. To cope with it, we use here
the von Roos form of $T$, containing three ambiguity parameters $\alpha$, $\beta$,
$\gamma$, constrained by the condition $\alpha + \beta + \gamma = -1$~\cite{vonroos}. Such
a form has the advantage of containing all the proposals made in the literature as special cases,
while having an inbuilt Hermiticity. Therefore the general form of the three-dimensional PDM
Schr\"odinger equation reads
\begin{equation}
  \left\{- \half \left[M^{\alpha}(\xb) \partial_i M^{\beta}(\xb) \partial_i
  M^{\gamma}(\xb) + M^{\gamma}(\xb) \partial_i M^{\beta}(\xb) \partial_i
  M^{\alpha}(\xb)\right] + V(\xb)\right\} \psi(\xb) = E \psi(\xb),  \label{eq:schrodinger} 
\end{equation}
where $\xb \equiv (x_1, x_2, x_3) = (x, y, z)$, $\partial_i \equiv \partial/\partial x_i$, $i=1$, 2,
3, $M(\xb)$ is the dimensionless form of the mass function $m(\xb) = m_0 M(\xb)$, $V(\xb)$ is
the potential and we have chosen units wherein $\hbar = 2 m_0 = 1$.\par
%
%
We can get rid of the ambiguity parameters in the kinetic energy operator by transferring them to
the effective potential energy $V_{\rm eff}(\xb)$ of the variable mass system~\cite{cq06,
bagchi04}. Equation (\ref{eq:schrodinger}) then acquires the form
\begin{equation}
  H \psi(\xb) \equiv \left(- \partial_i \frac{1}{M(\xb)} \partial_i + V_{\rm eff}(\xb)\right)
  \psi(\xb) = E \psi(\xb),  \label{eq:schrodinger-bis}
\end{equation}
where the explicit expression of $V_{\rm eff}(\xb)$ in terms of $V(\xb)$, $M(\xb)$, $\alpha$
and $\beta$ is given by 
\begin{equation}
  V_{\rm eff}(\xb) = V(\xb) + \half (\beta+1) \frac{\Delta M}{M^2} - [\alpha
  (\alpha+\beta+1) + \beta + 1] \frac{(\partial_i M)(\partial_i M)}{M^3}. \label{eq:Veff}
\end{equation}
\par
%
%
Here we shall consider the case where the mass only depends on $x$ and presents a solitonic
profile
\begin{equation}
  M(x) = \sech^2 qx,
\end{equation}
which has proved useful in various contexts~\cite{cq06, bagchi04, bagchi05}. Equation
(\ref{eq:Veff}) then reduces to
\begin{equation}
  V_{\rm eff}(\xb) = V(\xb) - 2q^2 [2\alpha (\alpha + \beta + 1) + \beta + 1] \cosh^2 qx + q^2
  [4\alpha (\alpha + \beta + 1) + \beta + 1].
\end{equation}
It may be interesting to observe that $V_{\rm eff}(\xb)$ coincides with $V(\xb)$ for the choice
of ambiguity parameters $\alpha = 0$, $\beta = -1$, first proposed by BenDaniel and
Duke~\cite{bendaniel} and sometimes advocated from first principles~\cite{levy}. However, since
plausible arguments have been put forward in favour of some other choices too (see,
e.g.,~\cite{ganguly} for recent reviews in one dimension), we shall not impose here any
constraint on the ambiguity parameters apart from that mentioned in the beginning.\par
%
%
\section{PDM model in a semi-infinite parallelepipedal channel}

To start with, let us consider an effective potential of the form
\begin{equation}
  V_{\rm eff}(\xb) = V_{\rm eff, 1}(x) + V_{\rm eff, 2}(y) + V_{\rm eff, 3}(z),
  \label{eq:V-parallel}
\end{equation}
where
\begin{equation}
   \begin{array}{lll}
      V_{\rm eff, 1}(x) & = - q^2 \cosh^2 qx + q^2 k(k-1) \csch^2 qx, & \qquad 0 < x < \infty,
          \\[0.4cm]
      & = + \infty, & \qquad - \infty < x < 0,  \label{eq:V1-parallel}
  \end{array}
\end{equation}
and 
\begin{equation}
  \begin{array} {lll}
       V_{{\rm eff}, i}(x_i) & = 0, & \qquad - \fraction < x_i < \fraction, \\[0.4cm]
       & = + \infty, & \qquad x_i < - \fraction \mbox{\ or\ }x_i > \fraction, 
\end{array}
\end{equation}
for $i=2$ and 3. In (\ref{eq:V1-parallel}), $k$ is assumed to be some positive constant. The
Schr\"odinger equation (\ref{eq:schrodinger-bis}) therefore amounts to
\begin{eqnarray}
  && \left[- \partial_x \cosh^2 qx \partial_x - \cosh^2 qx \left(\partial^2_y +
       \partial^2_z\right) - q^2 \cosh^2 qx + q^2 k(k-1) \csch^2 qx\right] \psi(x,y,z) \nonumber
       \\
  &&  \quad \mbox{} = E \psi(x,y,z)  \label{eq:eq-parallel} 
\end{eqnarray}
on the semi-infinite parallelepipedal domain $0 < x < \infty$, $-\pi/(2q) < y, z < \pi/(2q)$,
with the Dirichlet boundary conditions $\psi(0,y,z) = 0$, $\psi(x, \pm \pi/(2q),z) = 0$, $\psi(x, y,
\pm \pi/(2q)) = 0$.\par
%
%
The corresponding Hamiltonian $H$ commutes with the operators $L = - \partial^2_y$ and $M
= - \partial^2_z$. Their simultaneous normalizable eigenfunctions $\psi_{n,l,m}(x,y,z)$, fulfilling
the boundary conditions, can be obtained by separation of variable in (\ref{eq:eq-parallel}) and
are given by
\begin{equation}
  \psi_{n,l,m}(x,y,z) = \phi_{n,l,m}(x) \chi_l(y) \zeta_m(z),
\end{equation}
where 
\begin{equation}
  \chi_l(y) = \left\{\begin{array}{ll}
      \sqrt{\frac{2q}{\pi}} \cos[(l+1)qy] & \qquad {\rm for\ } l = 0, 2, 4, \ldots, \\[0.2cm]
      \sqrt{\frac{2q}{\pi}} \sin[(l+1)qy] & \qquad {\rm for\ } l = 1, 3, 5, \ldots,
  \end{array} \right.  \label{eq:chi}
\end{equation}
$\zeta_m(z)$ assumes a similar form with $m$ and $z$ substituted for $l$ and $y$, respectively,
and~\cite{gradshteyn}
\begin{equation}
  \phi_{n,l,m}(x) = {\cal N}_{n,l,m} (\tanh qx)^k (\sech qx)^{1+\delta}
       P^{\left(k- 1/2, \delta\right)}_n(1 - 2 \tanh^2 qx),  \label{eq:phi-parallel}
\end{equation}
with $P^{\left(k- 1/2, \delta\right)}_n(1 - 2 \tanh^2 qx)$ denoting a Jacobi polynomial,
\begin{eqnarray}
  \delta & = & \sqrt{(l+1)^2 + (m+1)^2}, \\
  {\cal N}_{n,l,m} & = & \left(\frac{2q \left(2n + k + \frac{1}{2} + \delta\right) n!\, \Gamma\left(n
       + k + \frac{1}{2} + \delta\right)}{\Gamma(n+1+\delta) \Gamma\left(n + k +
       \frac{1}{2}\right)}\right)^{1/2}.
\end{eqnarray}
Observe that the last functions $\phi_{n,l,m}(x)$ can be derived either directly or from the
known wavefunctions $\xi^{(\kappa, \lambda)}_n(z)$ of the constant-mass two-parameter
trigonometric P\"oschl-Teller potential $\kappa(\kappa-1) \csc^2 z + \lambda(\lambda-1)
\sec^2 z$, $0 < z < \pi/2$, via changes of variable $\exp(qx) = \tan[(1/2)(z + \pi/2)]$ and of
function $\phi_{n,l,m}[x(z)] = \sqrt{\cos z}\, \xi^{(\kappa, \lambda)}_n(z)$.\par  
\par
%
%
The simultaneous eigenvalues of $L$, $M$ and $H$ are $(l+1)^2 q^2$, $(m+1)^2 q^2$ and
\begin{equation}
  E_{n,l,m} = q^2 (2n+1+\delta) (2n+2k+\delta),  \label{eq:E-parallel}
\end{equation}
where $n$, $l$, $m=0$, 1, 2,~\ldots. The energy spectrum is therefore quadratic in $n$, with
some twofold degeneracies connected with the $(l, m)$ exchange, i.e., $E_{n,l,m} = E_{n,m,l}$
for $l \ne m$, as well as with some `accidental' degeneracies (i.e., degeneracies not related  to
any symmetry), such as $E_{n,1,8} = E_{n,5,6}$ corresponding to $\delta = \sqrt{85}$.
Note that the number of bound states is infinite.\par
%
%
\subsection{PDM model in a semi-infinite cylindrical channel}

Instead of (\ref{eq:V-parallel}), let us take
\begin{equation}
  V_{\rm eff}(\xb) = V_{\rm eff, 1}(x) + V_{\rm eff, 23}(\rho),
\end{equation}
where $V_{\rm eff, 1}(x)$ is still given by (\ref{eq:V1-parallel}), while
\begin{equation}
  \begin{array} {lll}
       V_{\rm eff, 23}(\rho) & = 0, & \qquad 0 \le \rho < R, \\[0.4cm]
       & = + \infty, & \qquad \rho > R, 
\end{array}
\end{equation}
with $\rho = \sqrt{y^2 + z^2}$. The Schr\"odinger equation now reads 
\begin{eqnarray}
  && \left[- \partial_x \cosh^2 qx \partial_x - \cosh^2 qx \left(\partial^2_{\rho} +
      \frac{1}{\rho} \partial_{\rho} + \frac{1}{\rho^2} \partial^2_{\varphi}\right) - q^2
      \cosh^2 qx + q^2 k(k-1) \csch^2 qx\right] \nonumber \\
  && \quad \mbox{} \times  \psi(x,\rho,\varphi) = E \psi(x,\rho,\varphi) 
\end{eqnarray}
on the semi-infinite cylindrical domain $0 < x < \infty$, $0 \le \rho < R$, $0 \le \varphi
< 2\pi$, with the new boundary conditions $\psi(0,\rho,\varphi) = 0$, $\psi(x,R,\varphi) =
0$, $\psi(x,\rho,2\pi) = \psi(x,\rho,0)$.\par
%
%
The two operators commuting with $H$ are now $L = - (\partial^2_{\rho} + \rho^{-1}
\partial_{\rho} + \rho^{-2} \partial^2_{\varphi})$ and $M = - {\rm i} \partial_{\varphi}.$
The simultaneous normalizable eigenfunctions of $H$, $L$ and $M$ can be written as
\begin{equation}
  \psi_{n,m,s}(x,\rho,\varphi) = \phi_{n,|m|,s}(x) \chi_{|m|,s}(\rho) \zeta_m(\varphi).
\end{equation}
Here
\begin{equation}
  \zeta_m(\varphi) = \frac{1}{\sqrt{2\pi}} e^{{\rm i} m\varphi}
\end{equation}
corresponds to the eigenvalues $m=0, \pm 1, \pm 2, \ldots$ of $M$. Furthermore,
\begin{equation}
  \chi_{|m|,s}(\rho) = {\cal N}_{|m|,s} J_{|m|}(\kappa_{|m|,s} \rho), \qquad \kappa_{|m|,s} =
  \frac{j_{|m|,s}}{R},
\end{equation}
where $J_{|m|}(z)$ is a Bessel function, the symbol $j_{|m|,s}$, $s=1$, 2,~\ldots, conventionally
denotes~\cite{abramowitz} its real, positive zeros, and~\cite{gradshteyn}
\begin{equation}
  {\cal N}_{|m|,s} = \sqrt{2} \left[R J_{|m|+1}(j_{|m|,s})\right]^{-1},
\end{equation}
provides normalized solutions to the eigenvalue equation
\begin{equation}
  L \chi_{|m|,s}(\rho) \zeta_m(\varphi) = \kappa_{|m|,s}^2 \chi_{|m|,s}(\rho) \zeta_m(\varphi), 
\end{equation}
which satisfy the second boundary condition. Finally, $\phi_{n,|m|,s}(x)$ and the energy
eigenvalues $E_{n,|m|,s}$ are still given by the right-hand sides of Eqs.~(\ref{eq:phi-parallel}) and
(\ref{eq:E-parallel}), but  with $\delta$ now defined by
\begin{equation}
  \delta = \frac{\kappa_{|m|,s}}{q} = \frac{j_{|m|,s}}{qR}.
\end{equation}
This time the only degeneracy of the energy spectrum is that connected with the sign of $m$.\par
%
%
\section{Conclusion}

In this Letter, we considered a particle with a PDM, moving in two different three-dimensional
quantum channels with an abrupt termination under the influence of some hyperbolic potential.
We showed that the lack of uniformity in the environment generates bound states, actually an
infinite number of them. This interesting physical property may be compared with a similar
phenomenon ocurring in quantum wires in the presence of a dot or a bend~\cite{olendski}.\par
%
%

\end{document}